 \documentclass[twocol]{ametsoc_tw}
 
 \pdfoutput=1
 \usepackage{amsmath} 
\usepackage{amssymb} 
\usepackage{mathrsfs} 
\usepackage{psfrag}
 
\newcommand{\red}[1]{{\color{red} #1}}
\newcommand{\blue}[1]{{\color{blue} #1}}
\newcommand{\be}[0]{\begin{equation}}
\newcommand{\ee}[0]{\end{equation}}
\renewcommand{\d}[2]{\frac{d #1}{d #2}} 

\newcommand{\id}{\mathrm{d}}

\newcommand{\ri}{\rho_i}
\newcommand{\rw}{\rho_w}
\newcommand{\ra}{\rho_a}
\newcommand{\vi}{\vec{v}_i}
\newcommand{\vw}{\vec{v}_w}
\newcommand{\va}{\vec{v}_a}

\newcommand{\dv}{\Delta\vec{v}}
\newcommand{\kv}{\hat{k}}

\newcommand{\faw}{\gamma}

\newcommand{\U}{\Lambda}
\newcommand{\half}{\tfrac{1}{2}}
\newcommand{\ab}{\frac{\alpha}{\beta}}

\newcommand{\ca}{\tilde{C}_a}
\newcommand{\cw}{\tilde{C}_w}
\journal{jcli}

\bibpunct{(}{)}{;}{a}{}{,}

\title{An analytical model of iceberg drift}

\authors{Till J.W. Wagner\correspondingauthor{Scripps Institution of Oceanography, University of California San Diego, 9500 Gilman Drive, La Jolla, CA 92093.}, Rebecca W. Dell, and Ian Eisenman}
\affiliation{University of California San Diego, La Jolla, California}
\email{tjwagner@ucsd.edu}
\abstract{
Iceberg drift and decay and the associated freshwater release are increasingly seen as important processes in Earth's climate system, yet a detailed understanding of their dynamics has remained elusive. Here, an idealized model of iceberg drift is presented. The model is designed to include the most salient physical processes that determine iceberg motion while remaining sufficiently simple to facilitate physical insight into iceberg drift dynamics. We derive an analytical solution of the model, which helps build understanding and also enables the rapid computation of large numbers of iceberg trajectories. The long-standing empirical rule of thumb that icebergs drift at 2\% of the wind velocity, relative to the ocean current, is derived here from physical first principles, and it is shown that this relation only holds in the limit of strong winds or small icebergs, which approximately applies for typical icebergs in the Arctic. It is demonstrated that the opposite limit of weak winds or large icebergs approximately applies for typical Antarctic tabular icebergs, and that in this case the icebergs simply move with the ocean surface current. It is furthermore found that when winds are strong, wind drag drives icebergs in the direction the wind blows, whereas weak winds drive icebergs at a 90$^\circ$ angle to the wind direction. 
}

\begin{document}

\maketitle
\section{Introduction} \label{sec:intro}
 
Recent years have seen an increased interest in the fate of icebergs shed from high-latitude glaciers. They remain a threat to shipping as well as offshore oil and gas exploration efforts. This is of particular relevance as retreating Arctic sea ice and increasing hydrocarbon demands have garnered the attention of industrial developers interested in both shipping and drilling in the Arctic Ocean \citep{Pizzolato:2014cy, NationalEnergyBoardCanada:UUcm8XfS}. 

Concurrently, ongoing global climate change is being held responsible for an observed increase in calving fluxes from Antarctic and Greenland glaciers, an increase that is projected to accelerate during the coming decades \citep[e.g.,][]{Rignot:2006fm, Copland:2007du, Rignot:2011hi,Joughin:2014ew} and that is expected to impact regional ecosystems and oceanographic conditions \citep[e.g.,][]{Vernet:2012th,Smith:2013cu,Stern:2015bo,Duprat:2016hw}. 

Furthermore, rapid shedding of icebergs from Northern Hemisphere ice sheets during the Heinrich Events of the last glacial period are believed to have affected oceanic and atmospheric conditions on a global scale \citep[see reviews in][]{Hemming:2004in,Stokes:2015dt}.

Motivated by factors such as these, icebergs have recently begun to be implemented in state-of-the-art GCMs \citep[e.g.,][]{Hunke:2011fx, Stern:2016kh}. An improved physical understanding of iceberg dynamics is crucial for this model development and will aid in the interpretation of the GCM simulation results.

Previous iceberg drift studies have often focused on the ability to \emph{(i)} reproduce individual iceberg trajectories using comprehensive dynamic hindcast models \citep{Smith:1983tk,Lichey:2001ww,Keghouche:2009wl} or  \emph{(ii)} predict trajectories using statistical relationships derived from observed trajectories. A well-known feature of the latter approach is the so-called 2\%-rule, which states that icebergs move at approximately 2\% of the wind velocity, relative to the ocean currents \citep[e.g.,][]{Garrett:1985ec,Smith:1993vi}. Other studies have focused on iceberg decay processes and the associated freshwater release into the high-latitude oceans \citep[e.g.,][]{Bigg:1997bp,Death:2006do,Martin:2010kb, Jongma:2013hz,Roberts:2014ff}. These studies typically use a representation of iceberg drift that is based on the model of \cite{Bigg:1997bp}.

Here, we examine the most salient characteristics of how iceberg trajectories are determined. 
We develop an idealized iceberg drift model which allows an analytical solution of iceberg velocities for given water and air surface velocities. As the iceberg trajectories are found to depend on iceberg size, we couple the drift model to an idealized decay model similar to \cite{Bigg:1997bp}. The Lagrangian iceberg model presented here is computationally inexpensive and requires only three input fields to simulate iceberg trajectories, namely ocean and atmosphere surface velocities and sea surface temperature (SST). Furthermore, the model's idealized form facilitates detailed physical interpretation and can therefore help build understanding of the physical processes that determine iceberg drift.

This article is structured as follows: Section \ref{sec:model} introduces the iceberg drift and decay representations, and the analytical drift solution. Section \ref{sec:ecco} presents  iceberg trajectories that are computed from the analytical solution, with surface conditions taken from an observational estimate. Section \ref{sec:limits} discusses the roles wind and currents play in determining iceberg trajectories, focusing on the limits of small icebergs (Arctic) and large icebergs (Antarctic).  Concluding remarks are given in Section \ref{sec:conc}.

\section{Iceberg drift model} \label{sec:model}

\subsection{Governing equation for iceberg drift}

We develop an iceberg drift model that is adapted from the canonical family of drift models used by \cite{Bigg:1997bp}, \cite{Gladstone:2001cq}, \cite{Martin:2010kb}, \cite{Roberts:2014ff}, and other studies. These models mostly differ in only minor details, and the momentum equation is typically written in the form
\be
M \d{\vi}{t} = -M f \kv \times \vi + F_w + F_a + F_p + F_r + F_i,
\ee
where $M$ is the mass of the iceberg, $\vi$ the iceberg velocity, and $f$ the Coriolis parameter. The terms on the right hand side represent the Coriolis force ($M f \kv \times \vi$), water drag ($F_w$), air drag ($F_a$), pressure gradient force ($F_p$), wave radiation force ($F_r$), and sea ice drag ($F_i$).

  The model developed in this study retains the main components of previous formulations, as discussed below. It is, however, somewhat idealized, with the central assumptions being as follows:
 \begin{enumerate}
 \item{Instantaneous acceleration is much smaller than other terms in the momentum balance, such that $M\id \vi/\id t \approx 0$ in equation (1).}
 \item{The pressure gradient force can be approximated from the ocean velocity by assuming geostrophy.} 
 \item{Iceberg velocity $\vi$ is much smaller than surface air velocity $\va$.}
 \item{Drag from sea ice and wave radiation are small compared to water drag, air drag, the Coriolis force, and the pressure gradient force.}
 \item{Water drag is dominated by the surface current $\vw$, such that vertical variations in the current over the depth of the iceberg can be ignored. Similarly, the wind felt by the iceberg at any height is assumed to be equal to the surface wind.}
 \end{enumerate}
The justification for each assumption is given below.

Assumption 1: This is best satisfied for small icebergs. Using the model simulations of Section \ref{sec:ecco} (below), we find that this assumption is typically fairly well satisfied for Arctic icebergs. When the iceberg length is less than 1.5\,km, we find that $M\id \vi/\id t$ estimated from the simulated velocities is typically less than 10\% of the air drag term. For very large icebergs such as those in the Antarctic, this assumption begins to break down: for lengths of about 20\,km, the iceberg momentum is typically of the same magnitude as the air drag term. However, comparing simulations of such large icebergs with and without applying assumption 1 shows close correspondence for iceberg trajectories (Fig.~S1) and meltwater distributions (not shown).

Assumption 2: the pressure gradient force has been argued to be well approximated by assuming a geostrophic ocean ocean velocity \citep{Smith:1983tk, Gladstone:2001cq, Stern:2016kh}. It should be noted that \cite{Bigg:1997bp} found a large ageostrophic component of the ocean velocity in regions of strong horizontally sheared flows, where this assumption may thus introduce substantial errors. 

Assumption 3: While the typical velocity scale of surface winds is $V_a \sim 10$\,m/s, icebergs tend to travel at speeds $V_i \sim 0.1$\,m/s \citep{Robe:2012wb}, such that typically $V_a \gg V_i$. 

Assumption 4: Previous studies have modeled wave radiation either explicitly \citep[e.g.,][]{Bigg:1996bz, Bigg:1997bp,Death:2006do,Jongma:2009cl} or included it in the wind drag term where it alters the drag coefficient \citep{Smith:1993vi,Keghouche:2009wl}.
However, a number of studies have found that the contribution of wave radiation is typically small compared to the air and water drag terms \citep{Bigg:1997bp,Gladstone:2001cq}. Here, we will therefore neglect this term.

Assumption 5 is invoked for simplicity as done in previous studies \citep[e.g.,][]{Martin:2010kb}.

 \begin{figure}[!t]
 \begin{center}
 \hspace{-.5 cm} \includegraphics[width=.8\linewidth]{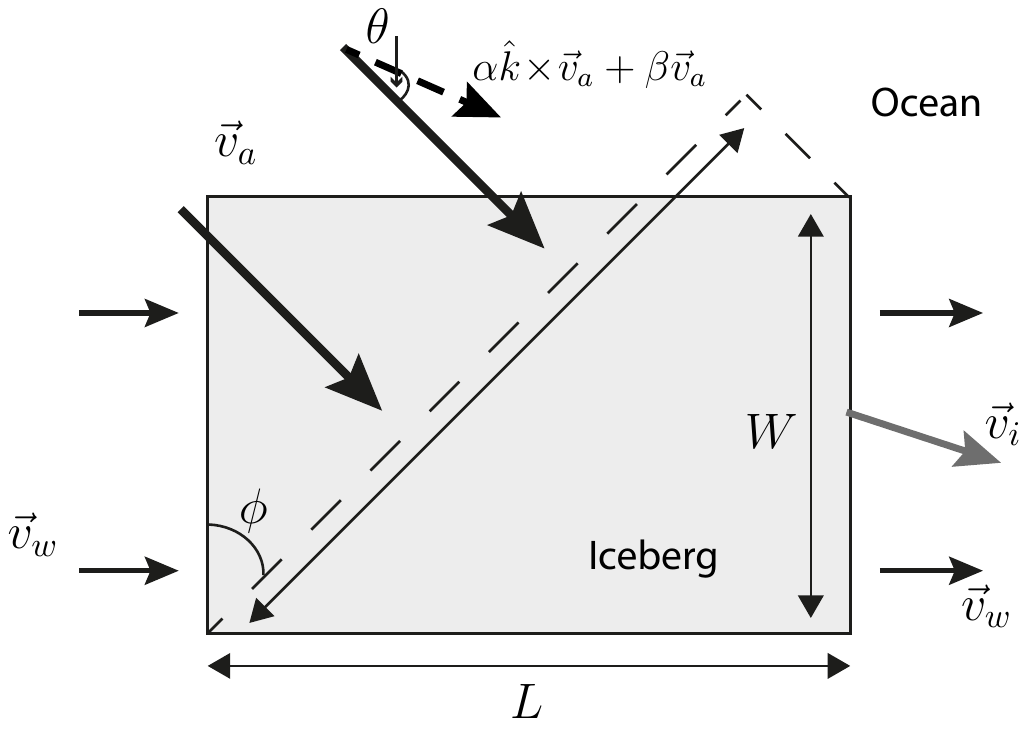}
 \caption{2-D bird's-eye view schematic of horizontal surface air and water velocities driving an iceberg (black arrows), as well as the iceberg velocity (gray arrow). Also indicated is the wind-driven component of the iceberg velocity, $\alpha\kv\!\times\!\va + \beta \va$, which acts at an angle $\theta = \tan^{-1}(\alpha/\beta)$ to the wind (dashed arrow). In the model we take the orientation to be random; in this schematic the water drag is acting along the distance $W$ and the air drag is acting along $W\cos\phi+L\sin\phi$.}
 \label{fig:bergschem}
 \end{center}
\end{figure}

The water and air drag terms are given by
\begin{align}
F_w &= \cw |\vw - \vi|(\vw - \vi), \\
F_a &= \ca |\va - \vi|(\va - \vi), \nonumber
\end{align}
where $\cw \equiv \half \rw C_w A_w$ and $\ca \equiv \half \ra C_a A_a$. Here $C_w$ and $C_a$ are bulk drag coefficients of water and air, $\rw$ and $\ra$ are water and air densities, and $A_w$ and $A_a$ are the cross-sectional areas on which the water and air velocities act, respectively. We take the icebergs to be cuboids of length $L$, width $W$, and height $H$, and we only consider drag exerted on the vertical surfaces of the icebergs. Previous studies often assumed a fixed orientation of the iceberg in the flow, such as the long axis being aligned with the ocean current as well as at 45$^\circ$ to the wind \citep{Bigg:1997bp}. We instead adopt the arguably more accurate approximation that icebergs are oriented at a random angle ($\phi$) relative to the wind and currents. In this case, the longterm mean horizontal length of the vertical working surface area for both drag terms is 
$
\tfrac{2}{\pi} \int_0^{\pi/2} \left( W \cos \phi + L \sin \phi \right) \id \phi = \tfrac{2}{\pi}(L+W),
$
such that 
\be
A_w = \frac{\ri}{\rw} \frac{2}{\pi} (L+W)H, \quad A_a = \frac{(\rw-\ri)}{\ri} A_w. \nonumber
\ee

The pressure gradient force is defined as
$
F_p \equiv - \left(M/\rho_w\right) \nabla P,
$
where $P$ is the horizontal pressure due to sea surface slope.  
Assuming that the pressure gradient force acting on the icebergs arises only from the geostrophic component of ocean flow (assumption 2) allows us to write the pressure force as 
$
F_p = M f \kv \times \vw.
$
The first two terms in equation (1) can then be combined. 

Making use of assumptions 1, 3, 4, and 5 then leads to
\be 
0 = M f \kv \times \dv+ \cw |\dv| \dv + \ca |\va| \va, \label{eq:mtm}
\ee
where $\dv \equiv \vw-\vi$ and $\va - \vi \simeq \va$ (assumption 3). Here, $\vw$ and $\va$ are understood to be surface velocities (assumption 5). 
As noted by \cite{Gladstone:2001cq}, the first term on the right hand side of equation \eqref{eq:mtm}  describes the deviation of the iceberg motion from the geostrophic component of the ocean current.  Note that in the second term the water drag is computed from the full $\vw$ field (including its ageostrophic components). 

The importance of the terms in equation \eqref{eq:mtm} can be quantified by the dimensionless quantities
\be
\Lambda_w \equiv \frac{\cw |\dv|}{Mf} \quad \mbox{and} \quad \Lambda_a \equiv  \frac{\ca |\va|^2}{Mf |\dv|},
\ee
which describe the magnitudes of the water and air drag terms relative to the Coriolis term, respectively. Note that these quantities are analogous to the Ekman number for fluid flow.

Equation \eqref{eq:mtm} can then be simplified to
\be
0 = \kv \times \Delta \hat{v} + \Lambda_w \Delta \hat{v} + \Lambda_a \hat{v}_a, \label{eq:mtmcmp}
\ee
where we use the unit vector notation $\hat{v} \equiv \vec{v}/|\vec{v}|$.

 \begin{figure*}[ht!]
 \psfrag{1}[c]{\red{$B_{1}$}}
  \psfrag{2}[c]{\blue{$B_{10}$}}
 \begin{center}
 \includegraphics[width=.9\linewidth]{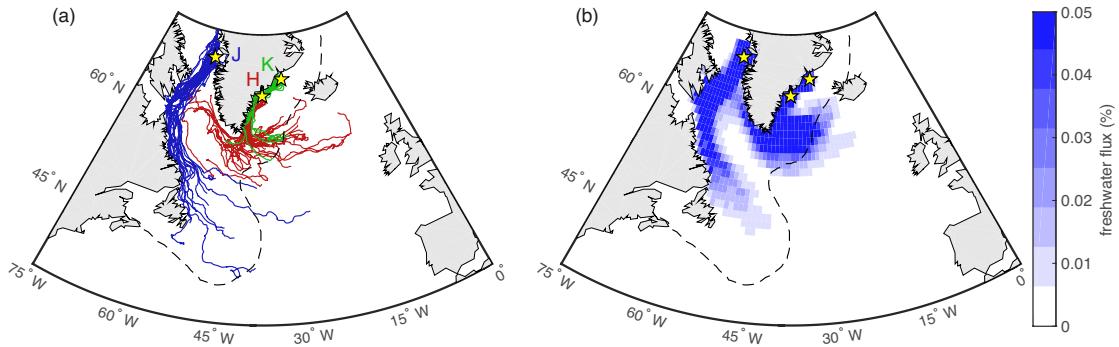} 
 \caption{Arctic iceberg trajectories. (a) Drift trajectories for icebergs released at three Greenland outlet glaciers. Shown are 50 track lines each for Kangerlussuaq (K, green), Helheim (H, red), and Jakobshavn (J, blue) glaciers. Also indicated are iceberg release locations (yellow stars), and as a dashed line the observed ``normal" range of icebergs as given by the \cite{IIP:xYIoaKr0}.  (b) Freshwater flux, computed from 1.5$\times10^4$ iceberg trajectories (500 for each of the 10 size classes from each of the three glaciers).}
 \label{fig:drift}
 \end{center}
\end{figure*}

\subsection{Analytical solution}

Equation \eqref{eq:mtmcmp} can be analytically solved for $\vi$, which is contained in $\Delta \hat{v}$, $\Lambda_w$, and $\Lambda_a$. Note that the vector equation \eqref{eq:mtmcmp} can be written as a set of two coupled scalar algebraic equations which are nonlinear in the components of $\vi$. The closed-form solution can be written as
\be
\vi =\vw + \gamma(\alpha \kv \times \va + \beta \va). \label{eq:sol}
\ee
Here $\gamma$ is a dimensionless parameter which describes the relative importance of water versus air drag,
\be
\faw \equiv \sqrt{\frac{\ca}{\cw}} =  \left[\frac{\ra (\rw-\ri)}{\rw \ri} \frac{C_a}{C_w}\right]^{1/2}.
\label{eq:gam}
\ee
The dimensionless parameters $\alpha$ and $\beta$ are
\begin{align}
\alpha &\equiv \frac{1}{2 \U^{3}}\left(1-\sqrt{1+4 \U^4}\right), \label{eq:ab}\\
\beta &\equiv \frac{1}{\sqrt{2} \U^{3}}\left[\left(1+\U^4\right)\sqrt{1+4\U^4}-3\U^4-1 \right]^{1/2}, \nonumber
\end{align}
where 
\be
\U \equiv \sqrt{\Lambda_w \Lambda_a}=  \frac{\gamma\, C_w}{\pi f}\frac{|\va|}{S},  \label{eq:U} 
\ee	
with $S \equiv L\,W/(L+W)$, the harmonic mean horizontal length of the iceberg. Since the typical range for modern icebergs features an approximately constant Coriolis parameter, $f$, the variable $\U$ can be interpreted as the ratio of wind speed to iceberg size. 
From equation \eqref{eq:sol} we see that the wind drives icebergs at an angle $\theta \equiv \tan^{-1}\left(\alpha/\beta\right)$. How $\theta$ varies with $\U$ will be discussed further in Section \ref{sec:limits}. Note that the iceberg velocity is independent of iceberg height $H$ since the drag terms and the Coriolis term scale with linearly $H$.

\subsection{Iceberg decay model}
Since the coefficients $\alpha$ and $\beta$ depend on iceberg length $S$, which is a function of $L$ and $W$, iceberg motion under given winds and currents will also depend on $L$ and $W$, and will therefore be affected by the decay of the iceberg. Iceberg decay is taken into account and modeled using a modified version of the thermodynamic decay model of \cite{Bigg:1997bp}, which accounts for three main melt processes. These are (i) wind-driven wave erosion; (ii) turbulent basal melt; and (iii) side wall erosion from buoyant convection. The main difference between the present model and that of \cite{Bigg:1997bp} is that we use the iceberg rollover criterion from \cite{Burton:2012hp}, instead of the account by \cite{Weeks:1978vi} which we find to be erroneous. See details in the Appendix.

\section{Iceberg trajectories and model validation from ECCO2 output} \label{sec:ecco}

The model is validated using the NASA ECCO2 product, a global ocean state estimate of the period 1992-2012 that is obtained using satellite and in situ data in concert with an ocean general circulation model \citep{Menemenlis:2008ve}. The surface wind forcing is taken from the Japanese 25-year ReAnalysis \citep[JRA-25,][]{Onogi:2007fx}. For simplicity we idealize the icebergs to be passive Lagrangian particles. This allows for the efficient computation of large numbers of iceberg trajectories. 

 We consider two scenarios: 
\begin{itemize}
\item[\emph{a.}]{Small icebergs (length $L<1.5$ km) released from three main outlet glaciers in Greenland.}
\item[\emph{b.}]{Large tabular icebergs ($L>15$ km) released off the coast of the Antarctic Peninsula.} 
\end{itemize}

The ECCO2 data set, which is used for both scenarios, consists of output fields averaged over 3 day intervals. We compute iceberg trajectories using equation \eqref{eq:sol} coupled to the decay model described in Section 2c. This requires as input the ECCO2 surface water and wind velocities, as well as the sea surface temperature. The JRA-25 wind velocities, which are given on a 1$^\circ$ grid, are interpolated onto the 0.25$^\circ$ grid of the ocean fields.

We integrate the iceberg trajectories using a Forward Euler time-stepping scheme with 1 day temporal resolution. The 3-day ECCO2 fields are linearly interpolated from time interval centers onto a 1 day time resolution. Iceberg velocities are computed at each time step using currents and winds from the spatial grid box that is centered nearest to the iceberg location.

 \begin{figure}[h]
 \begin{center}
\includegraphics[width=.9\linewidth]{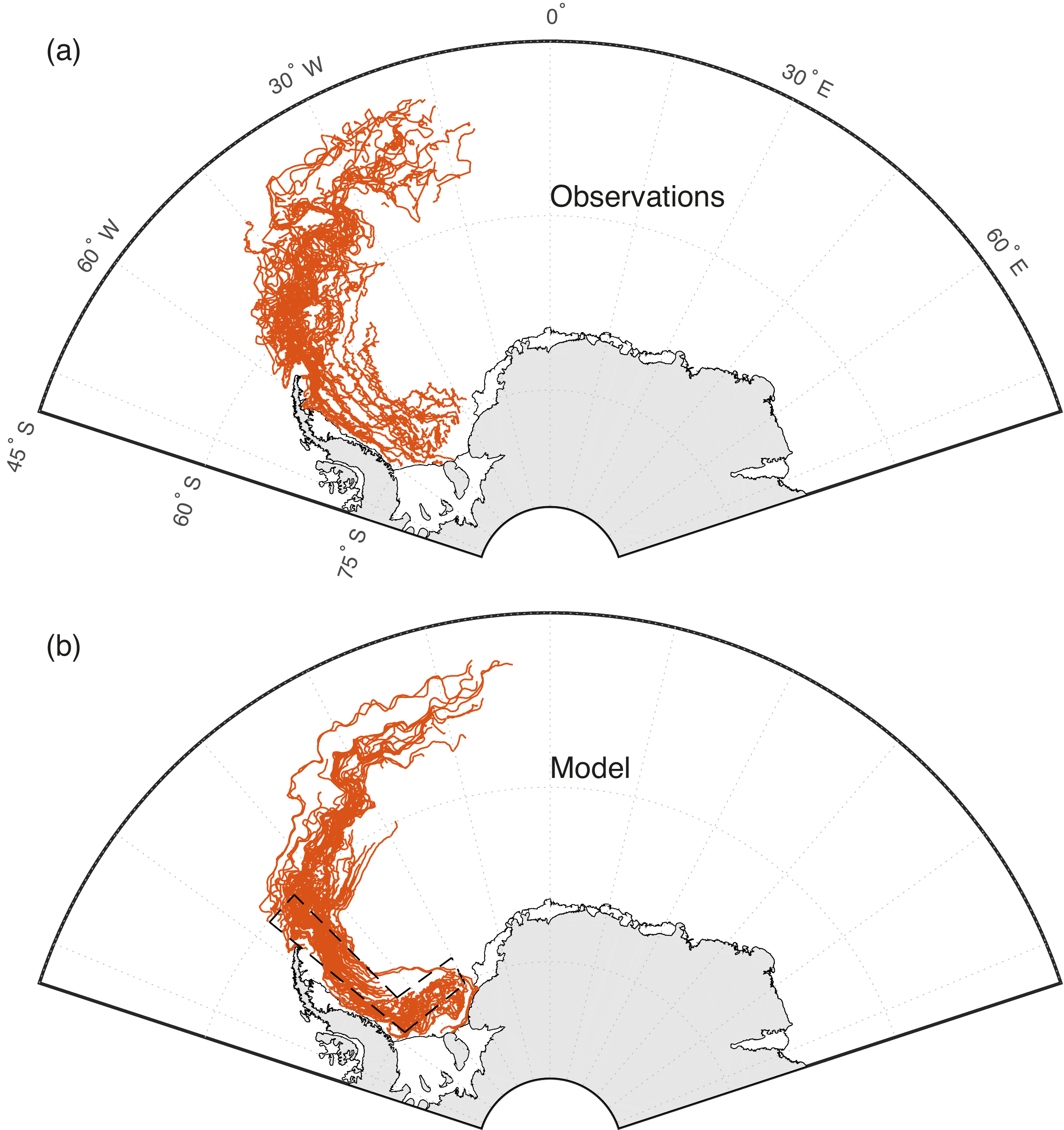}
 \caption{Antarctic iceberg trajectories. (a) Observed iceberg trajectories from the Antarctic Iceberg Tracking Database, which is derived from the QuickSCAT/SeaWinds scatterometer data (see text). Here the trajectories of icebergs with inception points around the Antarctic Peninsula are included.
(b) 200 modeled 1-year iceberg trajectories for large icebergs ($L = 15$-20 km), computed using equation \eqref{eq:sol} and ECCO2 input fields. The dashed polygon depicts the region of iceberg release locations in the model runs.}
 \label{fig:Ant}
 \end{center}
\end{figure}

Grounding events are not resolved explicitly; instead we set iceberg velocities to zero when icebergs come within one grid box of land until the surface circulation moves them away from the coast, akin to schemes used previously \citep{Wiersma:2010jj}. Matlab code for this iceberg drift model is available at http://eisenman.ucsd.edu/code. 

\subsection{Arctic iceberg simulations}
We release Arctic icebergs  near the outlets of Greenland's three main ice discharge glaciers \citep{Rignot:2006fm}: Helheim, Kangerlussuaq, and Jakobshavn (Fig.~\ref{fig:drift}). For each iceberg, the release location is chosen randomly from the centers of a $4 \times 4$ grid of ECCO2 grid points near each outlet glacier.
Icebergs are released at random times within the first 3 years (1992-1994) of the ECCO2 dataset and advected until they melt completely.
We consider 10 initial iceberg sizes, with dimensions ranging from $100 \times 67 \times 67$ m to $1500 \times 1000 \times 300$ m, following the classification of \citet[][their Table 1]{Bigg:1997bp}. A total of 500 icebergs are released for each size class from each glacier.

Fig.~\ref{fig:drift}a shows 50 iceberg trajectories for each of the three glaciers. Size classes and release dates for these trajectories are chosen at random. We find that most icebergs from Kangerlussuaq Glacier drift southward close to the coast, with frequent groundings. Small icebergs from Helheim Glacier are more commonly driven westward by winds, with a few larger Helheim icebergs drifting around the southern tip of Greenland (sizes are not indicated in Fig.~\ref{fig:drift}). Icebergs from Jakobshavn, on the other hand, quickly make their way across Baffin Bay and follow the ``Iceberg Alley" south along the Labrador Coast toward Newfoundland. They mostly melt completely by the time they reach the Grand Banks (approximately $45^\circ$N, $50^\circ$W). However, some simulated icebergs survive substantially longer and drift beyond the commonly observed iceberg boundary as estimated by the \cite{IIP:xYIoaKr0}, which is indicated by a dashed line in Fig.~\ref{fig:drift}. This may be partially due to the decay model not accounting for the breakup of large icebergs, which is likely a significant expediting factor in iceberg deterioration \citep[e.g.,][]{Wagner:2014uz}. Fig.~\ref{fig:drift}b shows the simulated freshwater input distribution due to iceberg melt. This is computed by averaging over all icebergs within one size class and weighing each iceberg size class field according to the log-normal distribution used in \citet[][their Table 1]{Bigg:1997bp} and subsequent studies. 

\subsection{Antarctic iceberg simulations}

We qualitatively validate the model against Antarctic tabular icebergs using the observed trajectories of large icebergs as catalogued in the Antarctic Iceberg Tracking Database. Fig.~\ref{fig:Ant}a shows the dataset of QuikSCAT/SeaWinds scatterometer observations \citep{Ballantyne:2002hn}, which tracked 352 icebergs of diameter $>10$ nm over the years 1999-2009.

Using the model \eqref{eq:sol}, we release 200 large icebergs of lengths between 15 and 20 km in the ECCO2 fields off the east coast of the Antarctic Peninsula and in the Weddell Sea, and we track these icebergs for 1 year (Fig.~\ref{fig:Ant}b). 
Note that we ignore the drag and reduced melting effects of sea ice in these simulations. The large uncertainties inherent in this comparison should be emphasized. Iceberg release locations, iceberg dimensions, and drift periods are among the unconstrained factors that make a direct comparison between model output and satellite observations difficult. Considering these uncertainties, the simulated trajectories show fairly good agreement with observations, accurately capturing the general drift pattern along the east coast of the Antarctic Peninsula and into the Antarctic Circumpolar Current.
It should be noted by caveat that the iceberg trajectories shown in Fig.~\ref{fig:Ant}b do not terminate with the disappearance of the icebergs, but rather with the end of the 1-year time window. Large tabular icebergs in these simulations would survive considerably longer, which is expected to mainly be an artifact of the model not accounting for breakup processes, as discussed above.

\section{The role of winds and currents} \label{sec:limits}

Here we address the roles that the three terms in equation \eqref{eq:sol} play in determining iceberg trajectories.
Specifically, we focus on the following questions:
\begin{itemize}
\item{Are icebergs primarily driven by winds or currents? To what degree does this depend on iceberg size and on the magnitudes of the air and water velocities?}
\item{What determines how much the wind drives icebergs along-wind versus across-wind? In other words, how does the angle $\theta$ depend on surface velocities and iceberg size?}
\end{itemize}

To address these questions, we consider first the analytical solution (Sec.~\ref{sec:limits}a) and subsequently the iceberg trajectories and velocities numerically computed from ECCO2 (Sec.~\ref{sec:limits}b).

\subsection{Winds and currents in the analytic model}

\begin{figure}[!ht]
 \begin{center}
 \includegraphics[width=.95\linewidth]{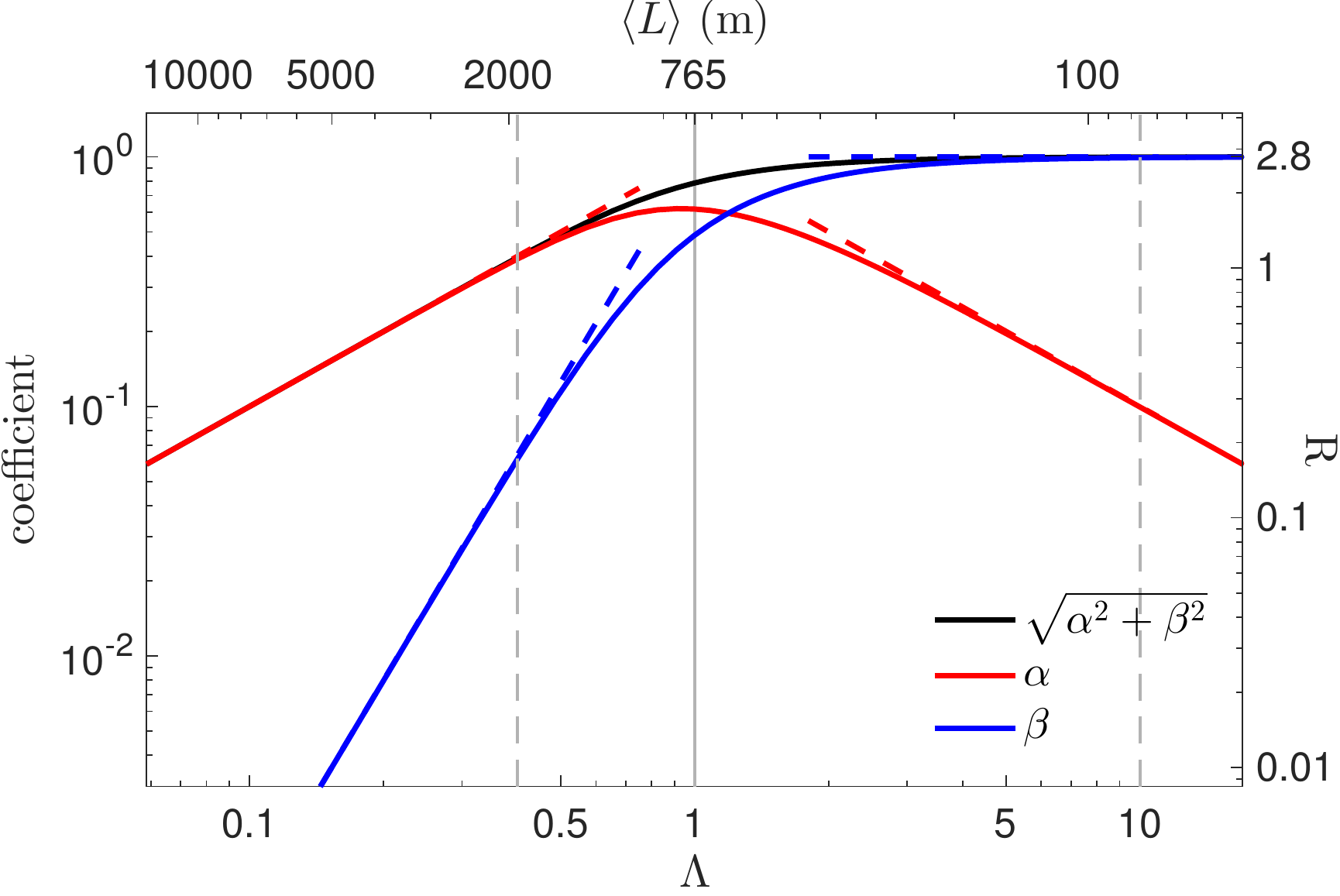}
 \caption{(a) Coefficients $\alpha$ and $\beta$, as well as $(\alpha^2 + \beta^2)^{1/2}$ in $\log$-$\log$ space. The asymptotics of equation \eqref{eq:abasy} are shown as dashed lines. The critical value $\U =  1$ (gray vertical line) and the regimes for $\alpha/\beta < 0.1$ and $\beta/\alpha <0.1$ (gray dashed lines) are also indicated.
The top horizontal axis shows the iceberg length $L $ for  $ |\va|  = 5.7$\,m/s (see text).
The right vertical axis shows the coefficient $R$, quantifying the relative importance of wind versus current forcing.} 
 \label{fig:ablog}
 \end{center}
\end{figure}

\paragraph{The direction of wind-driven motion.} The Coriolis term in equation \eqref{eq:mtm} causes some of the wind drag to project onto the direction perpendicular to the wind velocity, giving rise to the cross product term on the right hand side of equation \eqref{eq:sol}. The importance of this term relative to the along-wind term can be assessed by considering the coefficients $\alpha$ and $\beta$ (Fig.~\ref{fig:ablog}). We compute Taylor Series expansions of $\alpha$ and $\beta$ for small and large $\U$:
\be
 \alpha \simeq
\begin{cases}
\U \\
1/\U 
\end{cases}
\
\
 \beta \simeq
\begin{cases}
\U^3  &\ \mbox{for} \ \U \ll 1,\\
1  \ &\ \mbox{for} \ \U \gg 1. \label{eq:abasy}
\end{cases}
 \ee
 These asymptotics are also shown in Fig.~\ref{fig:ablog}.
 Note that, since $\U \propto |\va|/S$ is a measure of wind relative to iceberg size, one can think of these limits in two ways: $\U \gg 1$ applies to strong winds or, alternatively, small icebergs. $\U \ll 1$ applies to weak winds or large icebergs.

\begin{figure}[ht!]
 \begin{center}
\hspace{-.5cm}  \includegraphics[width=.95\linewidth]{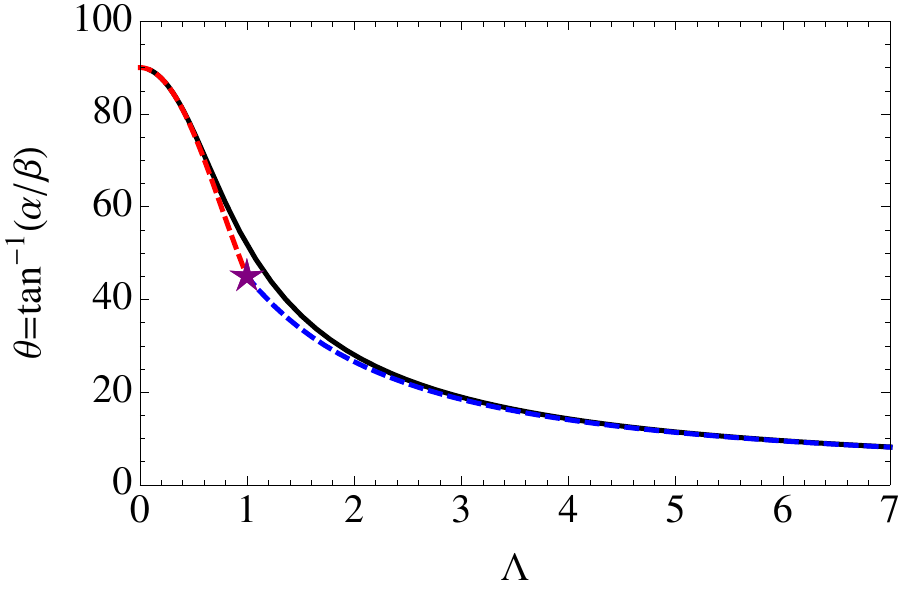}
 \caption{The direction in which the wind drives icebergs, $\theta$, as a function of dimensionless wind speed, $\U$ (solid). The dashed curves show the asymptotic approximations for small and large $\U$; see equation \eqref{eq:ablim}. These asymptotics meet at $\U=1 \ (\theta = 45^\circ)$, indicated by the purple star.}
 \label{fig:theta}
 \end{center}
\end{figure}

The angle, $\theta$, at which the surface winds drive the iceberg (relative to the direction of the wind) is indicated schematically in Fig.~\ref{fig:bergschem}, and its dependence on $\U$ is shown in Fig.~\ref{fig:theta}. From equation \eqref{eq:abasy} we obtain the asymptotic limits for the ratio $\alpha/\beta= \tan\theta$:
\be
\ab \simeq
\begin{cases}
\U^{-2}  &\ \mbox{for} \ \ \U \ll 1,\\
\U^{-1}  &\ \mbox{for} \ \ \U \gg 1.
\end{cases} \label{eq:ablim}
\ee
Both the angle $\theta$ and the ratio $\alpha/\beta$ decrease monotonically with increasing $\U$; i.e., the stronger the wind blows, the more it drives the iceberg in the direction of the wind. The two asymptotics meet when $\alpha = \beta$, in which case $\U= 1$ and $\theta=45^\circ$. Fig.~\ref{fig:theta} shows that these asymptotic solutions are a rather close approximation to the exact solution in the full range of $\U$, and we conclude that the wind drives the iceberg primarily along-wind when $\U > 1$ and across-wind when $\U<1$.

 \begin{figure*}[ht]
 \begin{center}
\hspace{1cm} \includegraphics[width=.7\linewidth]{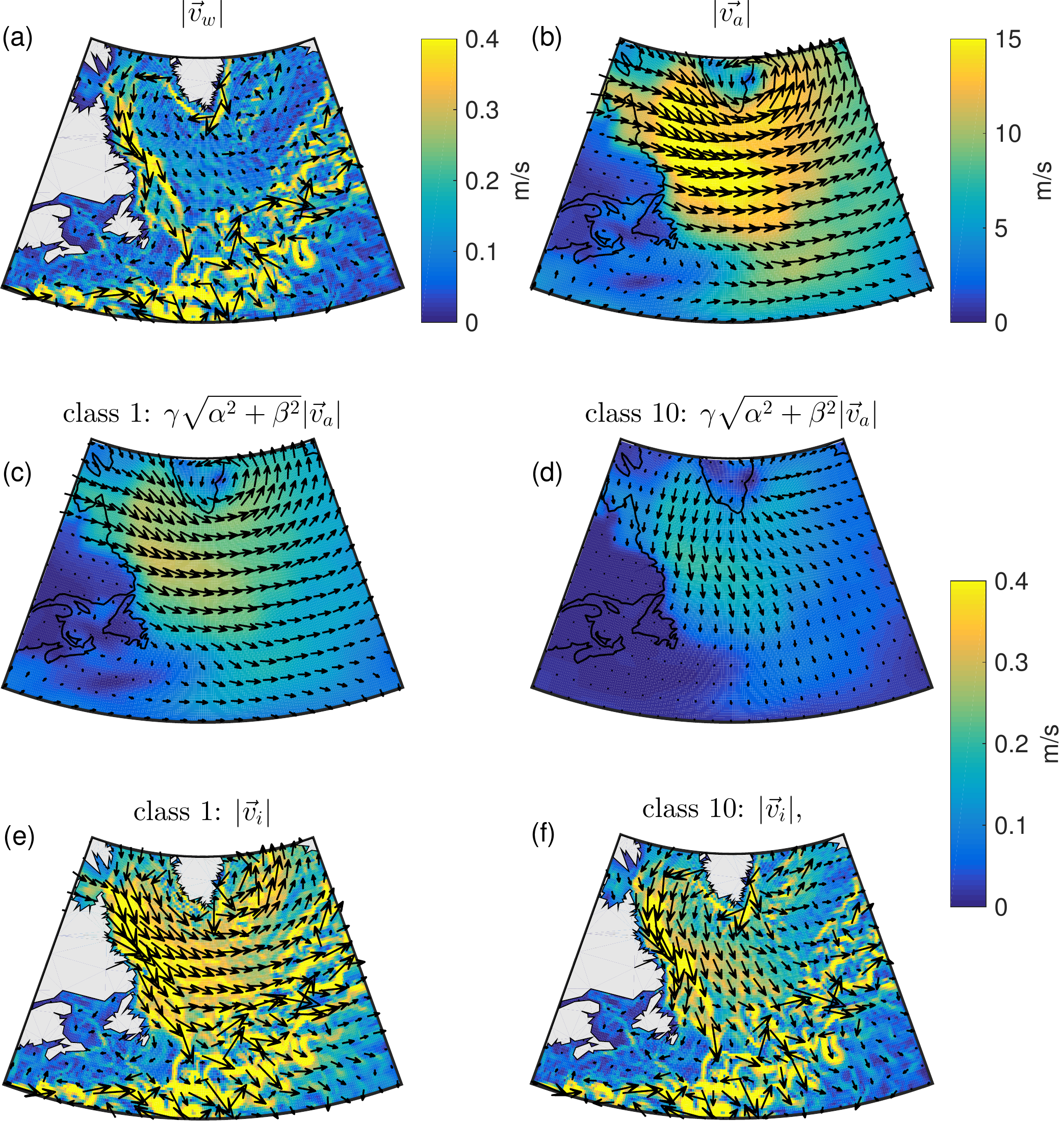}
 \caption{Velocities in the North Atlantic from ECCO2 output for 1-3 Jan 1993. (a) 3-day mean surface current speeds (shading) and velocity field (arrows). (b) Wind speeds and velocities for the same time period. (c) Wind-driven component of the iceberg velocity field for small (class 1) icebergs. (d) As in (c), but for large (class 10) icebergs. (e) Iceberg speed and velocity fields for small icebergs, corresponding to the sum of panels (a) and (c). (f) Iceberg speed and velocity fields for large icebergs, equal to the sum of (a) and (d).}
 \label{fig:vels}
 \end{center}
\end{figure*}

For strong winds, $\alpha/\beta \ll 1$ (or equivalently $\U \gg 1$), and $\theta \rightarrow 0$; i.e., the wind drives icebergs in the direction that it blows. For weak winds, on the other hand, $\alpha/\beta \gg 1$, and the wind drives icebergs at $\theta \simeq 90^\circ$ to the wind.
Next, we consider these limits in more detail.

\paragraph*{$\U \gg 1$ (strong winds, small icebergs).}
From equation \eqref{eq:abasy}, we find that for strong winds, or small icebergs, $\alpha \rightarrow 0$, and the along-wind component approaches a constant value, with $\beta \rightarrow 1$ (Fig.~\ref{fig:ablog}). Equation \eqref{eq:sol} therefore reduces to 
\be
\vi = \vw + \gamma \va.  \label{eq:sol2}
\ee
Using density values $\ra = 1.2$\,kg/m$^3$,  $\rw = 1027$\,kg/m$^3$, $\ri = 850$\,kg/m$^3$ and taking bulk coefficients $C_a =0.9$ and $C_w = 1.3$ \citep{Bigg:1997bp}, we find from equation \eqref{eq:gam} that $\gamma =  0.018$.

This is in close agreement with previous observational estimates, where icebergs have been empirically found to typically drift at about 2\% of the wind speed, relative to the water: the ``2\% rule''.  Such estimates include $\gamma = 0.018 \pm 0.7$ \citep{Garrett:1985ec} and $\gamma = 0.017$ \citep{Smith:1993vi}.

These results not only shed light on the origin of this empirical rule-of-thumb, but they also provide constraints for its validity: Icebergs satisfy the 2\% rule when $\U \gg 1$. We can interpret this dimensionally in terms of iceberg size. Assuming a constant Coriolis parameter $f=10^{-4}$s$^{-1}$ and a typical aspect ratio $L/W = 1.5$ \citep{Bigg:1997bp}, such that $S = 0.4 L$, gives 
\be
\U \simeq c |\va|/L, \label{eq:Uapprox}
\ee
where $c = 130$\,s. This means that the limit $\U \gg 1$ is satisfied when
$$
 |\va| \gg L/c.
$$
We conclude that the 2\% rule, which is often treated as a universal relationship, is a good approximation for iceberg drift \emph{only} when the surface air speed in m/s is much greater than ca.~1\% of the length of the iceberg in m. In Section 4b we will consider how this relates to the simulated icebergs. 

\paragraph*{$\U \ll 1$ (weak winds, large icebergs).}
In this limit, wind drives the icebergs predominantly in the direction perpendicular to the wind velocity. The reason for this is contained in equation \eqref{eq:sol}: the across-wind term is dominant for large icebergs since in this limit, the Coriolis term, which causes the across-wind component and scales with mass $M$, is large relative to the drag terms which scale with cross-sectional area.

 \paragraph{Surface winds versus currents.}
In the limit of small $\U$, both $\alpha$ and $\beta$ decrease when $\U$ decreases. This implies that the total wind contribution to iceberg motion drops off quickly for low winds or large icebergs. This is a further consequence of the Coriolis term growing large, since it depends only on $\vw$ (and not $\va$). Large tabular icebergs (as occur mostly in Antarctica) can therefore be assumed to be driven, to a first approximation, by the ocean currents alone, as we confirm in Section \ref{sec:limits}b below. The relative importance of wind and ocean currents can be quantified in terms of the ratio of the associated speeds. Previous studies have stated that the water drag either dominates \citep{Matsumoto:1996fv} or that water and air drags are of similar importance \citep{Gladstone:2001cq}. Here we aim to establish this relative importance quantitatively. We define the ratio of the magnitudes of the air and water velocity terms in equation \eqref{eq:sol} as
 \be
 R \equiv  \gamma (\alpha^2+\beta^2)^{1/2} |\va|/|\vw|. \label{eq:R}
 \ee
We will see below that the ratio $|\va|/|\vw|$ is approximately constant for icebergs from a given region. This implies that small icebergs (with $\alpha^2+\beta^2 \simeq 1$) move predominantly with the wind (i.e., $R>1$) when $|\va|/|\vw| > 1/\gamma \simeq 50$. For very large icebergs, the denominator in $R$ (i.e., the water velocities) will always be dominant, since in that case $\alpha^2+\beta^2 \rightarrow 0$. In the following we compare these limits to the actual velocity ratios experienced by icebergs simulated with the ECCO2 surface conditions.

\subsection{Wind, current, and ice velocities from ECCO2}

\paragraph{Arctic iceberg simulations.}A single 3-day snapshot (1-3 Jan 1993) of the ECCO2 velocities and corresponding magnitudes is shown in Fig.~\ref{fig:vels}. Also shown is the wind-driven component of iceberg velocities, $\gamma(\alpha \kv \times \va +\beta \va)$, and its magnitude, $\gamma (\alpha^2+\beta^2)^{1/2} |\va|$, for iceberg size classes 1 and 10 (second row). The third row of Fig.~\ref{fig:vels} represents the corresponding iceberg velocity fields.

In agreement with the discussion above, we find that (i) the wind-driven component is stronger for smaller icebergs and (ii) that its directionality is closely aligned with that of $\va$ for small icebergs, and more aligned with $\vw$ for large icebergs. 
In other words, the iceberg velocity field of the small size class is largely determined by the wind field, while the larger icebergs move primarily with $\vw$.

The relative importance of each term in equation \eqref{eq:sol} for the motion of Arctic icebergs can be quantified using the mean wind speeds experienced by the icebergs along their trajectory. These wind speeds are shown in Fig. S2. For a given glacier, the wind speed is found to be approximately constant, with the two East Greenland glaciers (Helheim, Kangerlussuaq) experiencing higher wind speeds near $6.5$ m/s and Jakobshavn having lower wind speeds near $4$ m/s. Inserting the average value of $|\va| = 5.7$ m/s for all Arctic icebergs into equation \eqref{eq:Uapprox} gives $L  = 765\mbox{m}\,/\U$. This length scale, indicated as the top horizontal axis of Fig.~\ref{fig:ablog}, acts as a measure of the relative importance of the wind terms in equation \eqref{eq:sol}. The critical length corresponding to $\U=1$, $L^* \equiv 765$m, separates the regimes where $\alpha$ and $\beta$ dominate. This means that Arctic icebergs will be driven along-wind if $L < L^*$ and mostly across-wind if $L > L^*$.

Analogously, the right vertical axis of Fig.~\ref{fig:ablog} shows the coefficient $R$ \eqref{eq:R} using the mean simulated velocity ratio $|\va|/|\vw|  = 150$. This means that the role of wind drag dominates that of water drag by a factor of $R\approx 3$ for small icebergs and that it becomes negligible compared to water drag ($R < 0.1$) for icebergs larger than $L \sim 12$ km. 

\begin{figure}[t!]
 \begin{center}
\includegraphics[width=.95\linewidth]{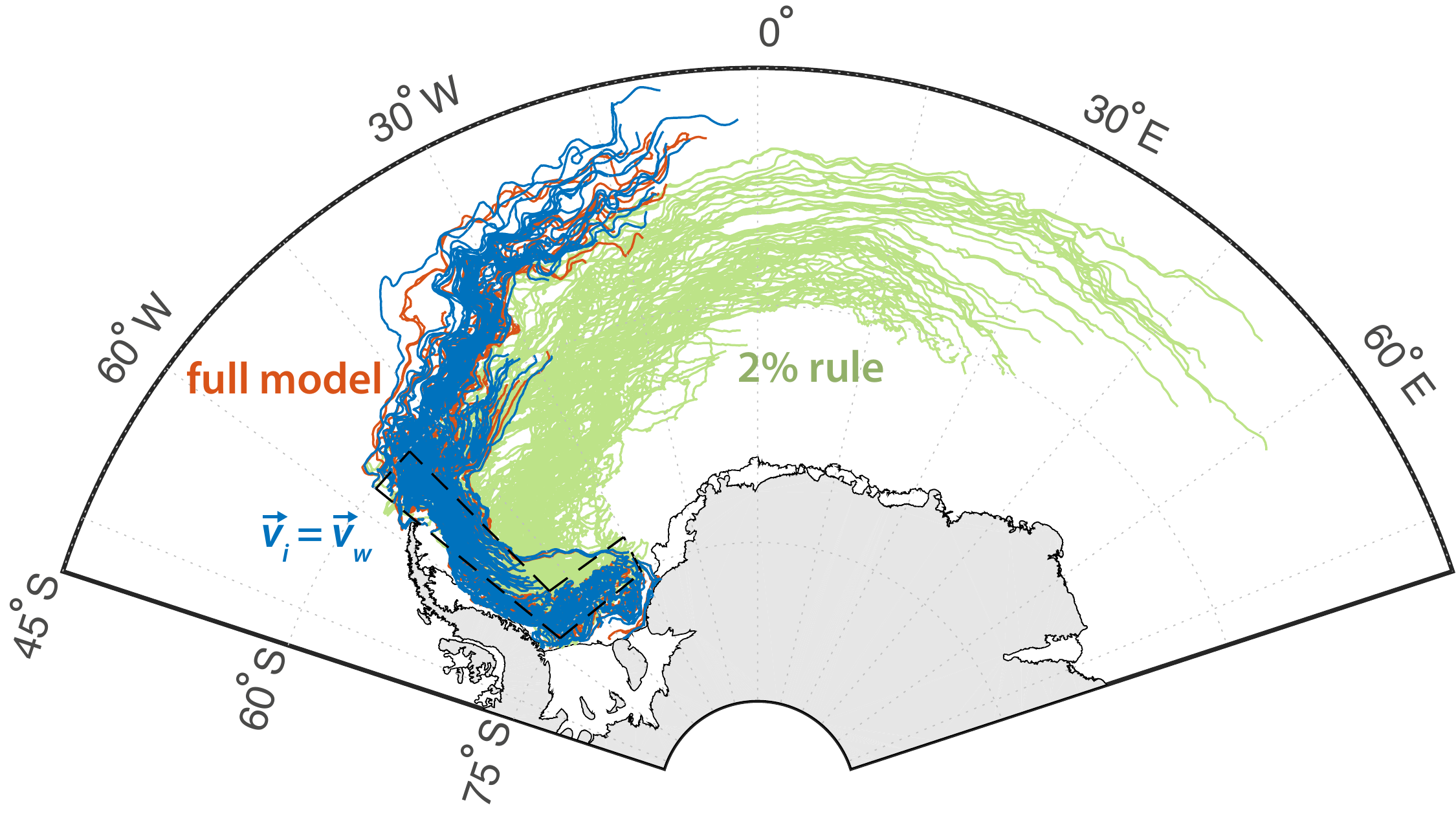}
 \caption{Simulated Antarctic iceberg trajectories. Shown are simulations of the full analytical model (red), iceberg motion that follows the ocean currents (blue, overlaying most of the red trajectories), and iceberg motion that follows the ``2\% rule'' (green). The dashed polygon depicts the region of iceberg release locations.}
 \label{fig:Ant2}
 \end{center}
\end{figure}

\paragraph{Antarctic iceberg simulations.}
The results above suggest that, for large tabular icebergs as observed in Antarctica, the wind drag can be assumed negligible and equation \eqref{eq:sol} reduces to the relation $\vi = \vw$, i.e., large icebergs move with the surface ocean current. To demonstrate the accuracy of this approximation, we perform two more sets of simulations using the same initial conditions as those for the Antarctic simulations discussed in Section \ref{sec:ecco}b (above). First, we approximate icebergs to move at the water velocity (blue trajectories in Fig.~\ref{fig:Ant2}). Next, we integrate the model with icebergs moving according to the 2\% rule (green trajectories in Fig.~\ref{fig:Ant2}). Whereas the icebergs with $\vi = \vw$ drift in close agreement with those using the full solution (red trajectories in Fig.~\ref{fig:Ant2}), the icebergs following the 2\% rule show a different drift pattern: they are substantially influenced by the strong prevailing winds around the Antarctic Peninsula. This implies that large icebergs move approximately with the ocean currents. 

\section{Conclusion} \label{sec:conc}

We have presented an idealized iceberg drift model with an analytical solution of iceberg motion for given surface circulation fields. This solution facilitates  
\begin{itemize}
\item{an improved understanding of the underlying mechanisms determining iceberg drift and}
\item{computationally inexpensive simulations of large numbers of iceberg trajectories.}
\end{itemize}
The model shows that in the limit of small icebergs and strong winds, icebergs drift at $\sim 2\%$ of the surface winds relative to the water. This asymptotic result of the analytical model is in good agreement with an empirical rule of thumb used in previous studies. However, the results highlight the limitations of this 2\% rule: in the limit of large icebergs and weak winds, the wind contribution to driving the icebergs becomes negligible, and the icebergs drift with the surface current. By considering trajectories computed from ECCO2 output fields, we find that these two limits approximately correspond to typical (small) Arctic icebergs and typical (large) Antarctic icebergs, respectively.

The dependence of the role of wind stress on iceberg size can be explained through the relative importance of the drag terms versus the Coriolis term: since the drag terms scale with surface area $LH$, and the Coriolis term scales with volume $L^2 H$ (assuming $L\sim W$), the latter will become large in the limit of large icebergs. And since the Coriolis term is independent of wind speed but depends on water velocity, water velocities dominate iceberg drift in this limit. 

Since iceberg calving rates from Greenland and Antarctica appear to be accelerating, iceberg dynamics are seen increasingly to be an important process in the climate system, and hence it is paramount to develop a more comprehensive understanding of how icebergs drift and decay.


\appendix

The three dominant melt processes, (i) wind-driven wave erosion $M_e$, (ii) turbulent basal melt $M_b$, and (iii) thermal side wall erosion from buoyant convection $M_v$, can be written as \citep{Bigg:1997bp}
\begin{eqnarray}
M_e & = & 0.5 S_s = 0.75 |\vec{v}_a - \vec{v}_w|^{0.5} + 0.05|\vec{v}_a - \vec{v}_w|, \nonumber\\ 
M_b & = & 0.58 \left| \vec{v}_w - \vec{v}_i\right|^{0.8}(T_w-T_i) L^{-0.2}. \label{eq:melt} \\
M_v & = & 0.0076 T_w + 0.0013 T_w^2,  \nonumber   
\end{eqnarray}
Other processses, such as surface melt, have been found to be small compared to these terms \citep{Savage:2001hz}.
The water temperature, $T_w$, is approximated by SST, and the iceberg temperature is assumed constant at $T_i =$ - 4$^\circ$C \citep{EL-Tahan:1987tk}. $S_s$ is the Douglas Sea State \citep{Martin:2010kb}. The melt terms in \eqref{eq:melt} are expressed in terms of meters per day of change in the iceberg's dimensions, with $T_w$ and $T_i$ in $^\circ$C and $L$ in m.

We further assume that these processes are linearly additive, such that iceberg volume evolves as
$dV/dt = d(LWH)/dt$,  with $dL/dt = dW/dt = M_e + M_v$, and $dH/dt= M_b$. 

As stated above, we replace the rollover condition of \cite{Weeks:1978vi} by that of \cite{Burton:2012hp}, who developed a potential energy minimizing condition to determine the point at which rectangular icebergs become unstable and tested this with laboratory experiments.  
Considering the aspect ration $\varepsilon \equiv W/H$, they found a critical value below which a rectangular iceberg is unconditionally unstable:  $\varepsilon_c = \sqrt{6\ri/\rw\left(1 - \ri/\rw \right)}$. Following \cite{Martin:2010kb}, we assume a uniform density for icebergs,  \mbox{$\ri = $ 850 kg m$^{-3}$}, which is based on tabular icebergs in the Southern Ocean. Taking water density to be \mbox{$\rw = $ 1025 kg m$^{-3}$}, gives \mbox{$\ri/\rw = $ 0.83} and \mbox{$\varepsilon_c = $ 0.92}. Note that \cite{Burton:2012hp} use \mbox{$\varepsilon_c = $ 0.75}.

As in previous studies, we assume that when $\varepsilon$ falls below $\varepsilon_c$, the iceberg capsizes. In the model, this means that we swap the shorter horizontal dimension of the iceberg $W$ for its height $H$.

We note that there are some apparent errors in the accounts of iceberg rolling in \cite{Weeks:1978vi} and several subsequent studies that adopted their representation. \cite{Weeks:1978vi} appear to have a sign error in their equation (9), where the last term should be positive. This sign error is adopted by \cite{Bigg:1997bp}, who in addition erroneously take icebergs to rotate along the long axis $L$ (not $W$). Both apparent errors are adopted by \cite{Jongma:2013hz} and \cite{Martin:2010kb}, with the latter study further replacing the iceberg height $H$ with the iceberg draft, $H \ri/\rw$. 

\bibliographystyle{ametsoc2014}
\bibliography{Drift}

%
%
%
%

\end{document}